\documentstyle[12pt]{article}

\newcommand{\eq}{\begin{equation}}
\newcommand{\en}{\end{equation}}
\newcommand{\eqa}{\begin{eqnarray}}
\newcommand{\ena}{\end{eqnarray}}

\newcommand{\lbl}{\label}

\begin{document}

\hskip 11cm \vbox{\hbox{DFTT 50/93}\hbox{September 1993}}
\vskip 0.4cm
\centerline{\bf   TWO-DIMENSIONAL QCD}
\centerline{\bf   ON THE SPHERE AND ON THE CYLINDER}
\vskip 1.3cm
\centerline{ M. Caselle, A. D'Adda, L. Magnea  and  S. Panzeri}
\vskip .6cm
\centerline{\sl Istituto Nazionale di Fisica Nucleare, Sezione di Torino}
\centerline{\sl  Dipartimento di Fisica
Teorica dell'Universit\`a di Torino}
\centerline{\sl via P.Giuria 1, I-10125 Torino, Italy}
\vskip 2.5cm

\begin{abstract}

The partition functions of QCD2 on simple surfaces admit representations
in terms of exponentials of the inverse coupling, that are modular
transforms of the usual character expansions. We review the construction
of such a representation in the case of the cylinder, and show how it
leads to a formulation of QCD2 as a $c=1$ matrix model of the
Kazakov-Migdal type. The eigenvalues describe the positions of $N$
Sutherland fermions on a circle, while their discretized momenta label
the representations in the corresponding character expansion. Using
this language, we derive some new results: we give an alternative
description of the Douglas-Kazakov phase transition on the sphere, and
we argue that an analogous phase transition exists on the cylinder. We
calculate the large $N$ limit of the partition function on the cylinder
with boundary conditions given by semicircular distributions of
eigenvalues, and we find an explicit expression for the large $N$ limit
of the Itzykson-Zuber integral with the same boundary conditions.
(Talk given at the ``Workshop on high energy physics and cosmology'' at
Trieste, July 1993.)
\end{abstract}
\vskip 3cm
\hrule
\vskip1cm
\noindent

\hbox{\vbox{\hbox{$^{\diamond}${\it email address:}}\hbox{}}
 \vbox{\hbox{ Decnet=(31890::CASELLE,DADDA,MAGNEA,PANZERI)}
\hbox{ internet=CASELLE(DADDA)(MAGNEA)(PANZERI)@TO.INFN.IT}}}
\vfill
\eject

\newpage

\section{Introduction}
In the last year, following the papers of Gross and Taylor \cite{GT},
there has been a renewed interest in the attempts to understand
two dimensional QCD in the large $N$ limit as a string theory.
In ref. \cite{GT} it was  shown that QCD2 is indeed a string theory by
proving that the coefficients of the expansion of the QCD2 partition
function in power series of $1/N$ can be interpreted in terms of
mappings from a two dimensional surface onto a two dimensional
target space.
A different approach was pursued in refs. \cite{MP,Dou,CDMP},
where the equivalence of QCD2 on a cylinder and on a torus with a
$c=1$ matrix model was shown in various ways.
In this talk we will describe the approach of \cite{CDMP},
which leads to expressions for the partition functions that are ``dual''
to the usual character expansions, in the sense that they can be
expanded in powers of the inverse coupling. This approach
is inspired by the interpretation of the Kazakov-Migdal model as a
model for high temperature QCD \cite{CAPht}, which we review in
Section 2. In Section 3 we will show, following ref \cite{CDMP},
how QCD2 on a cylinder and on a torus is equivalent
for any $N$ to a one-dimensional Kazakov-Migdal model, with the
eigenvalues of the matter fields costrained to live on a circle rather
than on a line. The generalization of these results to
Yang-Mills theories based on other classical
groups \cite{Pan} is also discussed, together with the equivalence
of these theories with the zero coupling limit of Sutherland
models \cite{Su}. In Section 4, we derive some
previously unpublished results concerning the large $N$ limit
of QCD2 on the sphere and on the cylinder.
We show how the phase transition on the sphere recently
discovered by Douglas and Kazakov \cite{DouKa}
can be understood in our language, and we exhibit its
close analogy with the phase transition of Gross-Witten \cite{GW} and
Wadia \cite{Wa}. Finally, applying the techniques of ref. \cite{Mat},
we argue that the partition function on the cylinder exhibits the same
transition\footnote{We have been informed \cite{Grosspriv} that the
existence of this transition has been independently established by D.
Gross and A. Matytsin.}.

\section{The Kazakov-Migdal model as high temperature lattice gauge
 theory}

Let us consider the Kazakov-Migdal model \cite{KM}
with quadratic potential defined on a $d$-dimensional
ipercubic lattice labeled by $x$:
\eq
S = \sum_{x} N {\rm Tr} \bigl[
 m^{2} \phi^{2}(x) - \sum_{\mu} \phi(x)U(x,x+\mu)\phi(x+\mu)
U^{\dagger}(x,x+\mu)\bigr] ,
\label{km}
\en
where $\phi(x)$ is an Hermitian $N \times N$ matrix defined on the sites
of the lattice, and $U(x,x+\mu)$ is a unitary $N \times N$ matrix,
defined  on the links, which plays the role of the gauge field as
in the usual lattice discretization of Yang-Mills theories.
The appearance of matter fields in the adjoint representation of the
gauge group suggests the possibility that they may be thought of as
the remnants of the gauge field along an extra compactified dimension,
in the limit where the compactification radius tends to zero.
Finite temperature lattice gauge theory in an example where this
situation is realized, with time as the extra compactified dimension,
and the temperature $T$ as the inverse of the compactification radius.
The corresponding Wilson action is given by
\eq
S_W=\frac{2N}{g^2}\sum_{n}~Re\left\{ T\sum_i~tr(U_{n;0i})
+\frac{1}{T}\sum_{i<j}~tr(U_{n;ij})\right\} ,
\label{wilson}
\en
where $ n \equiv (x,t)$ denotes the space-time position and
$U_{n;0i}$ ($U_{n;ij}$) are the time-like (space-like)
plaquette variables.

In the large $T$ limit the spacial plaquettes are of order $\frac{1}{T}$
and the timelike plaquettes are frozen around $U_{n;0i}=1$.
The action (\ref{wilson}) is invariant under a $Z_N$ symmetry
consisting in multiplying each timelike link by a space independent
element of the center of the group. The order parameter of such
symmetry is the Polyakov loop, defined by

\eq
P( x)= tr \prod_{t=1}^{N_t}(U_{ x,t;0}) .
\label{polya}
\en

At high temperature this symmetry is broken (deconfinement phase):
the timelike links $U_{x,t;0}$
and the Polyakov loop are frozen around one of the $Z_N$ invariant
vacua, for instance $1$,  hence $<P(x)> = 1$ .
In this situation it makes sense to expand $P(x)$ around its vacuum
expectation value and consider small fluctuations.
It was shown in ref. \cite{CAPht} that , in the large $T$ limit, if one
\begin{enumerate}
\item neglects the contributions coming from spacelike
plaquettes,
\item expands $P(x) = e^{i\frac{\phi (x)}{\sqrt{T}}}$
in powers of $\frac{1}{\sqrt{T}}$,
\end{enumerate}
one finds that the finite temperature lattice gauge theory in $d+1$
dimensions of eq. (\ref{wilson}) coincides, up to terms of order
$\frac{1}{T}$, with the
Kazakov-Migdal model of eq.(\ref{km}), in $d$ dimensions
and with $m^2=d$.

In the large $T$ limit the Wilson term corresponding to the spatial
plaquettes appears as a small (order
$\frac{1}{T}$) correction to the Kazakov-Migdal action. Something
similar
was proposed in ref. \cite{Dobro}, where a small Wilson term
is added to the Kazakov-Migdal model in order to break the local $Z_N$
symmetry and avoid superconfinement. The  results of ref. \cite{Dobro}
are in complete agreement with the large temperature
expansion proposed in
\cite{CAPht}: two phases are present, according to
whether the functional
integral is dominated by configurations with small
eigenvalues of $\Phi(x)$ (phase 1) or large eigenvalues of $\Phi(x)$
(phase 2).
In phase 1 the coefficient of the Wilson term remains small in the
continuum limit and the Wilson action can be consistenly treated as a
perturbation of the K-M action .
However, as a theory of induced QCD this fails to explain
the dominance of planar diagrams in the large $N$ limit .
Instead, from the point of wiew of the large $T$ expansion,
small eigenvalues of $\Phi(x)$ imply that  $<P(x)> = 1$.
We are in the high temperature
deconfining phase and the spatial plaquettes are consistently
depressed by the $\frac{1}{T}$ factor.
In this phase foam-like diagrams are expected to give a large
contribution and we have no planar diagram dominance at large $N$.
Quite the opposite happens in phase 2 where, according to \cite{Dobro},
the coefficient of the Wilson term is of order $N$ in the continuum
limit, so that the Wilson action is not any more a small perturbation
of the K-M term.
{}From our point of view, if the eigenvalues of $\Phi(x)$ are large,
the eigenvalues of $P(x)$ are distributed on the unit circle,
and  $<P(x)> =0$. The system is in the
confined phase and $T$ is below the critical temperature, hence the
spatial plaquettes contribution cannot be neglected. This is the
phase where the usual large $N$ behaviour is expected to hold.

In spite of this qualitative agreement with
the physical picture, the naive
large $T$ limit outlined above fails for $d>1$.
We know in fact that the K-M model with a quadratic potential
does not admit a continuum limit in that case \cite{Gross}.
On the other hand, it is known in the theory of finite temperature QCD
that the original picture \cite{Ap} of a complete dimensional
reduction does not work due to the infrared
divergences in the spacial plaquettes term, and one finds instead a three
dimensional gauge theory coupled with a scalar
field $\phi$ in the adjoint representation, with a non trivial potential
$V(\phi)$ (see for instance \cite{Re,po,weiss}).

The fundamental point for what follows is that none of these problems
occur in $d=1$: the continuum limit in the K-M model is possible with
quadratic potential at $m^2 = 1$, and there are no spatial plaquettes, so
that no approximation is done by neglecting them.
Besides, we know that two
dimensional QCD depends only on the area and the topology of the
surface on which it is defined, so at finite temperature it should
be independent of the temperature $T$.
Hence, we expect the infinite temperature limit
to give actually the {\it exact} result.
This is shown in the next section by working directly on
the functional integral of the continuum theory.

\section{QCD2 as a $c=1$ matrix model}

We shall study first the partition function ${\cal K}_2 ( g_1,g_2; t)$
of QCD2 on a cylinder, with fixed holonomies (Polyakov loops)
$g_1$ and $g_2$ at the two boundaries (say $x=0 $ and $x=2\pi$).
It is defined by the functional integral
\eq
{\cal K}_2 ( g_1,g_2; t)  =
\int {\cal D} A_\mu {\cal D} F
e^{-S(t)}
\hat{\delta} \left(W(0),g_1\right) \hat{\delta} \left(W(2 \pi),g_2\right)
\psi(g_1) \psi(g_2) ,
\label{eq4}
\en
where the action $S(t)$, as a result of the invariance of the theory
under area preserving diffeomorphisms, can be written as\footnote{The
coordinates on the cylinder are denoted by $x,
\tau$. Notice that we choose the time direction (with coordinate
$\tau$) as the compactified one.}
\eq
S(t) = \frac{2N}{t} \int_0^{2\pi} dx d\tau~ {\rm Tr} \{F^2 -
i F (\partial_0 A_1 - \partial_1 A_0) -iF[A_0,A_1] \} .
\label{eq2}
\en
Here $F$ and $A$ are independent fields and $t={\tilde g}^2 {\cal A}$,
with ${\cal A}$ the area. $W(x)$ denotes the Polyakov loop
\eq
W(x) = {\cal P} e^{i \int_0^{2\pi} d\tau A_0(x,\tau)} ,
\label{eq5}
\en
and $\hat{\delta}(g,h)$ denotes the conjugation invariant delta function
on the group manifold.
Finally  $ \psi(g_1) $ and $ \psi(g_2)$ are just normalization
factors that depend only on the eigenvalues of $g_1$ and $g_2$
and are chosen in such a way that the
sewing of two cylinders corresponds to a group integration.

In order to calculate the functional integral (\ref{eq4}) it is
convenient to choose the gauge $\partial_0 A_0 = 0$.
In this gauge all the non zero modes of the
Fourier expansion in $\tau$ of $A_0$ are set to zero.
In \cite{CDMP} it is shown that the corresponding Faddeev-Popov
determinant is exactly cancelled as a result of the integration
over the non zero modes of $A_1$ and $F$, so that in
the end  one obtains the following expression for ${\cal K}_2$:
\eqa
{\cal K}_2(g_{1},g_{2}; t) & = &
 \int {\cal D}B {\cal D}A e^{-\frac{N\pi}{t}
{\rm Tr} \int_{0}^{2\pi} dx \left[ \partial B - i[A,B] \right]^{2}}
\times \nonumber \\
& \times &
\hat{\delta} \left(W(0),g_1\right) \hat{\delta} \left(W(2 \pi),g_2\right)
\psi(g_1) \psi(g_2) .
\label{eq14}
\ena
where  $B(x)$ and $A(x)$ (matrix fields on the algebra)
denote the static modes of the $A_0(x,\tau)$ and $A_1(x,\tau)$ gauge
fields respectively. Eq (\ref{eq14}) is the first result of our analysis:
the dimensional reduction of the previous section is exact and
the result is, as expected, a KM model in one continuous dimension
(the spatial dimension of the cylinder), where  $A(x)$ and $B(x)$ play
respectively the role of  gauge field and matter field.
There is, however, one important difference between (\ref{eq14})
and the usual one-dimensional KM model: the boundary conditions depend
on $e^{2 \pi i B}$ rather than $B$.
This means that the eigenvalues of $B$ are defined on a circle
rather than on a line or, as explained in \cite{CDMP,Pan}, that
the matter field $B$ enters the
action through the unitary matrix defined by $e^{2 \pi i B}$.
At this point, by means of standard matrix model techniques, the
matrix $B(x)$ can be diagonalized, the functional integral over its
eigenvalues performed and  one finally obtains

\eq
{\cal K}_2(g_1,g_2; t)  =  \sum_P
       \frac{t^{-\frac{N-1}{2}}}{J(\theta) J(\phi)}
       \sum_{\{l_{i}\}}  (-1)^{P}
    \exp \left[ - \frac{N}{2t} \sum_{i = 1}^N
       \left( \phi_i - \theta_{P(i)} + 2 \pi l_i \right)^2     \right] .
\lbl{kcyl}
\en
where $\theta_i$ and $\phi_i$ are the invariant angles of $g_1$ and
$g_2$, respectively\footnote{For the $SU(N)$ group the invariant
angles $\theta_i$ and $\phi_i$ as well as the integers $l_i$ are
constrained by $\sum_i \theta_i = \sum_i \phi_i = \sum_i l_i = 0$.
These constraints are not there in the $U(N)$ case,
where however an extra factor $(-1)^{(N-1) \sum_i l_i}$
is present at the r.h.s.}, while $J(\sigma)$ is, up to a phase the
Vandermonde determinant for a unitary matrix,
\eq
J(\sigma) = \prod_{i<j} 2 \sin \frac{\sigma_i - \sigma_j}{2} .
\lbl{vandermonde}
\en

In (\ref{kcyl}) the normalization factor $\psi(g)$ has also been
determined. Up to a constant, it is given by
$\psi(g) = J(\theta)/\Delta(\theta)$, where $ \theta_i $ are the
invariant angles of $g$ and $\Delta(\theta)$ is the usual Vandermonde
determinant of a hermitian matrix.

The formula (\ref{kcyl}) is one of the main results of our analysis.
It provides for  ${\cal K}_2$ an expansion in exponentials of $1/t$,
which is related to the well known character expansion
by a non trivial modular transformation \cite{AltItz}, as explicitly
checked in~\cite{CDMP}.

By taking the $\theta_i \rightarrow 0$ limit of eq. (\ref{kcyl}),
one obtains the modular inversion for the kernel on the disk,
${\cal K}_1(\phi; t)$, which has been known for many years
\cite{MeOn},
\eqa
{\cal K}_{1}(\phi; t) & = & {\cal N}_1 (N)
\times t^{-\frac{N^2 - 1}{2}}
 \sum_{\{l_i\} = - \infty}^{+ \infty}
\prod_{i<j=1}^N \frac{\phi_i - \phi_j + 2 \pi (l_i - l_j)}{2 \sin
\frac{1}{2}\left[\phi_i - \phi_j + 2 \pi (l_i - l_j)\right]} \nonumber
\times \\
& \times &
\exp \left[-\frac{N}{2t} \sum_{i = 1}^N \left(\phi_i + 2 \pi l_i\right)^2
\right] .
\lbl{meon}
\ena

The kernel on the cylinder, as given by eq. (\ref{kcyl}), can be
interpreted as follows: let $\theta_i$ and $\phi_i$ be
respectively the initial and final positions of an ensemble
of $N$ particles on a circle, and
let $t$ be the time elapsed in the transition from the initial to the
final configuration.
Then ${\cal K}_2 (g_1,g_2; t)$ can be interpreted as the
transition probability for such process. Notice that, because of the
Vandermonde determinants, the wave functions are completely
antisymmetric under the exchange of pairs of these particles,
which must therefore be fermions.
That QCD2 would prove to be equivalent to a quantum theory of free
fermions on a circle was already to be expected from eq. (\ref{eq14}).
It is known in fact  that one dimensional KM model describes the
singlet sector of an $SU(N)$ matrix model
with $c=1$ \cite{KazBoul,CAP}, whose interpretation in terms of
free fermions has been known for some time.

The fermionic system that describes QCD2 turns out to be a particular
case of a family of integrable one dimensional quantum mechanical
models first investigated by Calogero \cite{Calo} and Sutherland
\cite{Su}. In fact, one can easily check from eq. (\ref{kcyl})
that ${\cal K}_2 (g_1,g_2; t)$ satisfies the differential equation
\eq
\left( N\frac{\partial}{\partial t} - \frac{1}{2}  J^{-1}(\phi)
\sum_i \frac{\partial^2}{\partial \phi_{i}^{2}}  J(\phi)
\right) {\cal K}_2 (\phi,\theta; t)  =  0 .
\label{heateq}
\en
By taking the limit $\theta_i \rightarrow 0 $ in eq. (\ref{heateq})
one finds that also the kernel on a disk ${\cal K}_1(\phi; t)$
obeys the same equation.
Upon redefinition of the kernel by ${\cal K}_1 \rightarrow J(\phi)
{\cal K}_1 \equiv \hat{\cal K}_1$, eq. (\ref{heateq}) becomes the
(euclidean) free Schr\"{o}dinger equation for $N$ fermions on a circle.
The boundary conditions are determined by
this redefinition, and are just the boundary conditions that apply to
the fermions in the zero coupling limit of the Sutherland
integrable model related to the Lie algebra $A_{N-1}$ \cite{Su}.

The same equation, but with $J(\phi)$ replaced by $\Delta(\phi)$,
is satisfied by the corresponding quantities in one dimensional
hermitian matrix models, where the place of ${\cal K}_2$ is taken
by the Itzykson-Zuber integral (see for instance the recent paper by
Matytsin \cite{Mat}). Such models are then related to the zero
coupling limit of the Calogero integrable model \cite{Calo},
in the same way as QCD2 is related to the Sutherland model.

All quantities of relevance to QCD2 can readily be reinterpreted
in the Sutherland language, as discussed in more detail
in \cite{Pan}. For example, we already mentioned that
${\cal K}_2 (\theta,\phi; t)$ can be interpreted as
the propagator from an initial configuration $\theta_i$ to a final
one $\phi_i$ in a time $t$. Similarly, the kernel on the disk
${\cal K}_1 (\phi; t)$ corresponds to the
transition amplitude from the configuration $\theta_i = 0$
to $\phi_i$, and the partition function on the sphere is just the
amplitude for the process in which all the fermions start at the origin
and come back there after a time $t$.
The meaning of the modular
transformation relating the expression
(\ref{kcyl}) to the usual character expansion is also clear:
the integers labelling the unitary representations
in the character expansion correspond to discretized momenta
of the fermions on
the circle,  while eq. (\ref{kcyl}) gives the corresponding coordinate
representation.

The partition function of QCD2 on a torus can be obtained from eq.
(\ref{kcyl}) by identifying $\theta_i$ with $\phi_i$ and then
integrating with the group
invariant measure $\prod_i d\theta_i J^2 (\theta)$.
The integral is gaussian in the invariant angles $\theta_i$,
although with a
complicated combinatorial structure due to the sum over all
permutation in eq. (\ref{kcyl}). This can be disentangled by
decomposing each permutation in cycles
and by calculating the integrals corresponding to cyclic permutations.
The result has a particularly simple form if one considers the grand
canonical partition function  at $\tilde{t} = \frac{t}{N}$ fixed.
For $SU(N)$ for instance one gets
\eqa
{\cal Z}_{G=1}^{SU(N)} (q)&  = & \sum_N {\cal Z}_{G=1}^{SU(N)}(N, \tilde{t})
q^N
\nonumber \\
& = & \left( \frac{\tilde{t}}{4\pi} \right)^{1/2} \int_{0}^{2
\pi}
d\beta \prod_{n=-\infty}^{+\infty} \left( 1 + q
e^{- \frac{\tilde{t}}{2} \left( n - \frac{\beta}{2\pi}
\right)^2 } \right) .
\label{gasfermionsSUN}
\ena

The interpretation of this formula as the grand canonical
partition function of a gas of free fermions with energy levels that
go like $n^2$ for $n \rightarrow \infty$ is evident.
The parameter $\beta$ is the momentum corresponding to the center
of mass of the fermions, which in $SU(N)$ is localized (see footnote
following eq. (\ref{kcyl})). Hence the corresponding momentum $\beta$ is
completely undetermined.

All the results obtained in this section can be generalized to the
case of an arbitrary simple gauge group \cite{Pan}.
One obtains that the matrix model
describing a two dimensional Yang-Mills theory on a cylinder is a KM
model where the matter fields are in the fundamental representation of
the gauge group. Since the string interpretation of this matrix model
requires expanding the group matrices as exponentials of the matrices on
the algebra, one has an orientable string theory for the $SU(N)$ group,
while if the gauge group is $SO(N)$ or $Sp(2N)$ the worldsheet of the
string may be both orientable or nonorientable \cite{SOSP}.

\section{Large N phase transitions on the sphere and on the cylinder}

In this section we report on some new results concerning
mainly the large $N$ limit of the kernel on the cylinder and the
partition function on the sphere, where a large $N$ phase transition
was recently found by Douglas and Kazakov \cite{DouKa}.
Unlike the authors of \cite{DouKa}, we work in configuration space
and consider the expression for the partition function on a sphere
of area ${\cal A}$, obtained by sewing disks of areas $t$ and
${\cal A} - t$,

\eqa
{\cal Z}_{G=0}(N, {\cal A}) & = & \int_0^{2\pi} d\theta_i J^2(\theta)
{\cal K}_1 \left( \theta, t \right)
{\cal K}_1 \left( \theta, {\cal A}-t \right) \nonumber \\
& = & \left( \frac{t({\cal A}-t)}{4N^2}\right)^{-\frac{N^2 -1}{2}}
\int_{-\infty}^{+\infty} d\lambda_i
\sum_{\{ n_i \}} \Delta(\lambda) \Delta(\lambda + 2 \pi n) \times
\nonumber \\
& \times & \exp \left\{ - N \sum_i \left[
\frac{\lambda_i^2}{2t} +\frac{ (\lambda_i +2\pi n_i)^2}{2({\cal A}
- t)} \right] \right\} ,
\label{matrmod}
\ena
where we used eq. (\ref{meon}) for ${\cal K_1}$ and, in the first
kernel, we replaced $\theta_i + 2 \pi m_i$ with a variable
$\lambda_i$ defined on the whole real axis. We want to find the
eigenvalue distribution that, for any given $t$, dominates the
integral in eq. (\ref{matrmod}) in the large $N$ limit.
The crucial point is that, if the eigenvalue distribution is, for any
$t$, confined in the interval $(-\pi,\pi)$, then we are entitled
to consider in eq. (\ref{matrmod}) only the term corresponding
to $n_i = 0$, for all $i$. The other terms in
fact describe the winding of the eigenvalues, which is possible,
at $N=\infty$, only if the eigenvalue distribution spreads at some
$t$ over the whole circle.
The term $n_i = 0$ describes a gaussian matrix model whose eigenvalue
distribution is given by Wigner's semicircle law
\eq
\rho(\lambda,  r) = \frac{2}{\pi r^2} \sqrt{r^2 - \lambda^2} ,
\label{Wigner}
\en
where the radius $r$ of the distribution is given in terms of $t$ and ${\cal
A}$
by
\eq
r = 2 \sqrt{\frac{t({\cal A}-t)}{{\cal A}}} .
\label{rad}
\en

The maximum value of the radius of the eigenvalue distribution occurs at
$t=\frac{{\cal A}}{2}$, and it is given by $r_{\max} = \sqrt{{\cal A}}$.
According to the previous argument, $r_{\max}$ must not exceed $\pi$,
which implies ${\cal A} < {\pi}^2$.
In complete agreement with the results of ref. \cite{DouKa} we find
then two phases: one for ${\cal A} < {\pi}^2$, where the partition
function on the sphere is completely described in the large $N$ limit
by a gaussian matrix model, and the winding modes can be neglected.
In this region the Sutherland and Calogero models are indistinguishable
at large $N$. The other phase, for ${\cal A} > {\pi}^2$, is
characterized in configuration space by winding numbers different from
zero. It is remarkable that, in configuration space,
the mechanism of the phase transition discovered by Douglas and
Kazakov is the same as the one of the Gross-Witten-Wadia phase
transition \cite{GW,Wa}.

This reasoning can be extended to the cylinder if one realizes that,
from the point of view of the dynamics of the eigenvalues, the sphere is
just a cylinder with special boundary conditions, namely that all
eigenvalues must be concentrated at the origin at the initial and final
times. To study the cylinder in the large $N$ limit, in the gaussian
phase, one needs to generalize eq. (\ref{rad}), to find the time
evolution of the radius of the semicircular eigenvalue distribution
for arbitrary boundary conditions (that is allowing for non-zero
initial and final radii). This can be done by making use of the results
of a recent paper by Matytsin \cite{Mat}, which studies the time
evolution of the density of eigenvalues in a system satisfying
eq.(\ref{heateq}), with $J(\phi)$ replaced by $\Delta(\phi)$.
The result is that the evolution is determined by the Das-Jevicki
hamiltonian \cite{DJev}, and the corresponding equations of motion
can be written as the Hopf equation
\eq
\frac{\partial f}{\partial t} +
f \frac{\partial}{\partial \lambda} f = 0 ,
\label{hopf}
\en
where $f(\lambda,t) \equiv v(\lambda,t) + i \pi \rho(\lambda,t)$,
and $\rho(\lambda,t) $ is the density of eigenvalues $\lambda_i$ at
time $t$.
According to our discussion, the replacement of $J(\phi)$ with
$\Delta(\phi)$ in eq. (\ref{heateq}) does not affect the large $N$
limit in the gaussian phase. Therefore, the semicircular distribution
(\ref{Wigner}), with radius given by (\ref{rad}),
must be a solution of (\ref{hopf}). Indeed, if one inserts in
(\ref{hopf}) the ansatz (\ref{Wigner}), with $r=r(t)$, one finds
that (\ref{hopf}) is solved if
\eq
\ddot{r} + \frac{4}{r^3} = 0 .
\label{errecubo}
\en
The general solution of this differential equation is easily found,
and it generalizes naturally eq. (\ref{rad}), as
\eq
r(t) = 2 \sqrt{\frac{(t+\alpha)(\beta - t)}{\alpha + \beta}} .
\label{gensol}
\en
Given arbitrary boundary conditions $r(0)= \gamma_0$ and
$r({\cal A})= \gamma_1$, one can find from (\ref{gensol}) the
corresponding values of $\alpha$ and $\beta$. In particular, the sphere
corresponds to $\gamma_0 = \gamma_1 = 0$, which implies $\alpha = 0$ and
$\beta = {\cal A}$, in agreement with (\ref{rad}).
Eq. (\ref{Wigner}), with $r$ given by (\ref{gensol}), gives then the
classical trajectory for the density of eigenvalues that dominates, in
the large $N$ limit, the functional integral corresponding to the
kernel ${\cal K}_2$, with area ${\cal A}$
and boundary conditions given by Wigner's distributions of radii
$\gamma_0$ and $\gamma_1$.
The actual value of ${\cal K}_2$ in the large $N$
limit is given by
\eq
{\cal K}_2 (\gamma_0,\gamma_1; {\cal A}) = e^{- N^2 S_{\rm class}} ,
\label{kappa2}
\en
where $S_{\rm class}$ is the action calculated on the classical
trajectory,
\eq
S_{\rm class} = \frac{1}{8} \int_0^{\cal A} dt
\left( \dot{r}^2 + \frac{4}{r^2} ,
\right)
\label{azcl}
\en
with $r$ given by (\ref{gensol}).

Now it is easy to see that the same mechanism that leads to a phase
transition on the sphere operates also on the cylinder.
The solution (\ref{kappa2}) is the
large $N$ limit of the QCD2 kernel on a cylinder only if
$r(t) < \pi$ for any $t$ on the trajectory.
For any given boundary conditions (that is for any given $\gamma_0$
and $\gamma_1$), the maximum value of $r(t)$ increases as the area
${\cal A}$ increases (this is because $\ddot{r} < 0$ from
(\ref{errecubo})), so it is
bound to hit the value $\pi$ at some critical value of ${\cal A}$,
which can easily be determined from the initial conditions.
There a phase transition occurs, in analogy with case of the sphere.

Our solution for the large $N$ limit of ${\cal K}_2$ of course applies
also to matrix models where the eigenvalues live on the real axis
rather than on a circle, but in that case with no restrictions on the
values of $r(t)$, and no phase transitions.
In other words, our solution applies without restrictions
if one is dealing with a Calogero rather than a Sutherland model.
In this case the solution (\ref{kappa2}) is equivalent to
the calculation of the Itzykson-Zuber integral in the large $N$ limit
\cite{Mat}, with Wigner's distribution of
the eigenvalues. We find
\eqa
\frac{1}{N^2} \ln I_{IZ}(\gamma_0,\gamma_1) & = &
- \frac{1}{2} \ln \left( \frac{1}{2} + \frac{1}{2}
\sqrt{1+ \frac{\gamma_0^2 \gamma_1^2}{4}} \right) \nonumber \\
 & - & 1/2 \left( 1 - \sqrt{1+
\frac{\gamma_0^2 \gamma_1^2}{4} }\right) + O(1/N) ,
\label{itzub}
\ena
where $I_{IZ}(\gamma_0,\gamma_1)$ is the Itzykson-Zuber integral,
calculated with respect to two sets of eigenvalues that in the
large $N$ limit have Wigner distributions of radii $\gamma_0$ and
$\gamma_1$. Notice that, for $\gamma_0 = \gamma_1$, the expression
(\ref{itzub}) reduces to the one calculated by Gross
in \cite{Gross}.

We see that the ``dual'' expansions for the partition functions of
QCD2, which have been determined so far for the disk, the sphere, the
cylinder and the torus, are valuable tools for the study of the theory,
both for finite $N$ and in the large $N$ limit. They provide a physical
insight which is complementary to the one given by the character
expansions, just as configuration space is complementary to momentum
space, and they have a very direct interpretation in terms of
Sutherland fermions, which leads to a simple understanding of the phase
transitions on the sphere and on the cylinder.

\end{document}